\documentclass[twocolumn,showpacs,preprintnumbers,amsmath,amssymb]{revtex4}

\usepackage{graphicx}
\usepackage{dcolumn}
\usepackage{bm}
\begin{document}

\title{Explanation of the Stern-Gerlach splitting of spinor
condensates based on symmetry}

\author{Zhibing Li and Chengguang Bao\footnote{corresponding author: stsbcg@zsu.edu.cn}}

\affiliation{The State Key Laboratory of Optoelectronic Materials
and Technologies \\ School of Physics and Engineering \\ Sun
Yat-Sen
University, Guangzhou, 510275, P.R. China \\ and \\
Center of Theoretical Nuclear Physics, National Laboratory of
Heavy Ion Collisions, Lanzhou 730000, P. R. China}

\begin{abstract}The Stern-Gerlach splitting of spinor
condensates is explained based on the total spin-states with
specified SU(2) and permutation symmetries.

\end{abstract}

\pacs{ 03.75.Fi, \ 03.65.Fd}

\maketitle

The experimental realization of the spinor Bose-Einstein
condensation in optical traps \cite{stam98,sten98,myat98} is a great
step in probing the microscopic world. In the experiment by Stenger,
et al. , \cite{sten98} after the atoms had been trapped sufficiently
long by the optical trap, the trap was suddenly switched off and the
atoms are allow to expand, and a magnetic field gradient was applied
to yield a Stern-Gerlach splitting. \ Then the expanding cloud was
observed via the absorption imaging where the atoms are divided into
three group according to the hyperfine component $\mu =1,\ 0,$ and
$-1$ . The number of atoms in each hyperfine state can be evaluated.
We found that there is a strong symmetry background governing the
splitting, this is reported as follows.

After a sufficient long time of evolution, the system of condensed
atoms would arrive at a status of equilibrium, and would be
distributed among the low-lying eigen-states, the probability
$P(E_{i})$ of staying at a
particular eigen-state $\Psi _{i}$ is determined by thermodynamics, namely, $%
P(E_{i})\propto $ $e^{-E_{i}/kT}$, where $T$ is the temperature. Let
$M$ be the $Z$-component of the total spin. During the evolution
$M$\ remains unchanged. Furthermore, when all $N$\ atoms condense
into the same spatial state, the spatial wave function must be
completely symmetric with respect to particle interchanges,
accordingly the total spin-state must also be completely symmetric.
Furthermore, due to the property of the interaction, the total spin
$S$\ together with $M$ are good quantum numbers for all the
eigenstates. \ Thus, to understand the orientation of the spins, it
is crucial to understand the completely symmetric normalized\ total
spin-states $\vartheta _{S,M}^{[N]},$ where $S$ is ranged from $N,
N-2,\cdots $\ to $M$ (or $M+1$) if $N-M$ is even (or odd)
\cite{katr01,bao04} .

When $N$\ is small, $\vartheta _{S,M}^{[N]}$ is simple, e.g., for $%
N=3,$
\begin{equation}
\vartheta _{1,M}^{[3]}=\frac{\sqrt{5}}{3}[(\chi (1)\chi
(2))_{0}\chi (3)]_{1,M}+\frac{2}{3}[(\chi (1)\chi (2))_{2}\chi
(3)]_{1,M} \label{eq1}
\end{equation}
where $\chi (i)$\ is the spin-state of the $i$-th particle,
particles $1$ and $2$ are first coupled to spin zero and two,
respectively, then all three particles are coupled to $S=1$.
However, when $N$ is larger, $\vartheta
_{S,M}^{[N]}$ becomes very complicated. Fortunately, the expression of $%
\vartheta _{S,M}^{[N]}$\ itself is not really necessary. \ Making
use of the fractional-parentage coefficients $a_{S}^{[N]}$ and
$b_{S}^{[N]}$ derived in our previous papers, \cite{bao05} we can
extract anyone of the particles (say, particle 1) from $\vartheta
_{S,M}^{[N]}$ as
\begin{eqnarray}
\vartheta _{S,M}^{\lbrack N\rbrack }&=&a_{S}^{\lbrack N\rbrack
}\lbrack \chi (1)\vartheta _{S+1}^{\lbrack N-1\rbrack }\rbrack
_{SM}+b_{S}^{\lbrack N\rbrack }\lbrack \chi (1)\vartheta
_{S-1}^{\lbrack N-1\rbrack }\rbrack _{SM} \nonumber \\
&=&
a_{S}^{[N]}\sum_{\mu }C_{1\mu ,\;S+1,M-\mu }^{S\;M}\;\chi _{\mu
}(1)\vartheta _{S+1,M-\mu }^{[N-1]} \nonumber \\
&+& b_{S}^{[N]}\sum_{\mu }C_{1\mu ,\;S-1,M-\mu }^{S\;M}\;\chi
_{\mu }(1)\vartheta _{S-1,M-\mu }^{[N-1]} \label{e2}
\end{eqnarray}
where
\begin{equation}
 a_{S}^{[N]}=[(1+(-1)^{N-S})(N-S)(S+1)/(2N(2S+1))]^{1/2}
 \label{e3}
 \end{equation}
\begin{equation}
b_{S}^{[N]}=[(1+(-1)^{N-S})\;S%
\;(N+S+1)/(2N(2S+1))]^{1/2} \label{e4}
\end{equation}
and $C\vspace{1pt}_{1\mu ,\;S\pm 1,M-\mu }^{S\;M}$ are the
Clebesh-Gorden coefficients. It is clear from (\ref{e3}) and
(\ref{e4}) that $N-S$ must be even.

For the state $\vartheta _{S,M}^{[N]}$, from (\ref{e2}), the
probability of a spin at $\mu $ is
\begin{equation}
P_{\mu }^{S,M}=(a_{S}^{[N]}\;C_{1\mu ,\;S+1,M-\mu
}^{S\;M})^{2}+(b_{S}^{[N]}\;C_{1\mu ,\;S-1,M-\mu
}^{S\;M})^{2}\label{e5}
\end{equation}
Not only the fractional parentage coefficients, the related
Clebesh-Gorden coefficients in (\ref{e5}) have also analytical
forms, \cite{edmo60} i.e.,
\begin{eqnarray}
&&C_{11,\;S+1,M-1}^{S\;M}=[\frac{%
(S-M+1)(S-M+2)}{(2S+2)(2S+3)}]^{1/2}  \\
&& C_{10,\;S+1,M}^{S\;M}=-\;\lbrack \frac{%
2\;(S+M+1)(S-M+1)}{(2S+2)(2S+3)}\rbrack ^{1/2} \\
&& C_{1,-1,\;S+1,M+1}^{S\;M}=\lbrack \frac{%
(S+M+1)(S+M+2)}{(2S+2)(2S+3)}\rbrack ^{1/2} \\ &&
C_{11,\;S-1,M-1}^{S\;M}=\lbrack \frac{%
(S+M-1)(S+M)}{2S(2S-1)}\rbrack ^{1/2}\\
&&
C_{10,\;S-1,M}^{S\;M}=\;\lbrack \frac{%
2\;(S+M)(S-M)}{2S(2S-1)}\rbrack ^{1/2} \\ &&
C_{1,-1,\;S-1,M+1}^{S\;M}=\lbrack \frac{%
(S-M-1)(S-M)}{2S(2S-1)}\rbrack ^{1/2} \label{e6}
\end{eqnarray}
Thus $P^{S,M}_{\mu}$ has an analytical form as
\begin{eqnarray}
P_{1}^{S,M}&=&\frac{1}{2(2S+1)}(\frac{(1-S/N)(S-M+1)(S-M+2)}{2S+3} \nonumber \\& &+%
\frac{(1+(S+1)/N)(S+M-1)(S+M)}{2S-1}) \label{e7a}\\
P_{0}^{S,M}&=&\frac{1}{(2S+1)}(\frac{(1-S/N)(S+M+1)(S-M+1)%
}{2S+3} \nonumber \\ & &+\frac{(1+(S+1)/N)(S-M)(S+M)}{2S-1}) \label{e7b}\\
P_{-1}^{S,M}&=&P_{1}^{S,-M} \label{e7c}
\end{eqnarray}
and in general
\begin{equation}
P_{\mu }^{S,-M}=P_{-\mu }^{S,M} \label{e8}
\end{equation}

It is recalled that $\vartheta _{S,M}^{[N]}$\ are completely
symmetric, therefore the spin of each particle has exactly the same
probability $P_{\mu }^{S,M}$. From (\ref{e7a}-\ref{e7c}) we have
\begin{eqnarray}
&& P_{1}^{S,M}+P_{0}^{S,M}+P_{-1}^{S,M}\equiv 1  \label{e9}\\
&& N\;(P_{1}^{S,M}-P_{-1}^{S,M})\equiv M \label{e10} \\
&&\frac{1}{2S+1}\overset{S}{\underset{M=\;-S}{\sum }}%
P_{\mu }^{S,M}=\frac{1}{3} \label{e11}
\end{eqnarray}

Eq.(\ref{e9}) is a basic requirement because $\mu $\ has only three
choices, (\ref{e10}) implies that $NP_{\mu }^{S,M}$ is the number of
bosons at $\mu $. \ Eq.(\ref{e11}) implies that, for a \
nonpolarized system, the probability of a particle staying at a
given $\mu $ is $1/3$.

In particular, when $M=0$, we have
\begin{eqnarray}
&& P_{1}^{S,0}=P_{-1}^{S,0}=\frac{(1-1/2N)S(S+1)-1}{(2S+3)(2S-1)}%
\label{e12} \\
&&P_{0}^{S,0}=\frac{(2+1/N)S(S+1)-1}{(2S+3)(2S-1)} \label{e12b}
\end{eqnarray}

Since $N$ is usually large, we can neglect the term \ $1/N$. \ Then
we found both $P_{1}^{S,0}$ and $P_{-1}^{S,0}$ are close to $1/4$
and $P_{0}^{S,0}$ is close to $1/2$ (unless $S$ is very small). This
is a crucial point to explain the splitting experiment in
Ref.\cite{sten98}.

When $M=N$ (in this case $S=N$ is the only choice) we have $%
P_{1}^{N,N}=1,\ P_{0}^{N,N}=P_{-1}^{N,N}=0$ as expected.

\begin{figure}
\includegraphics{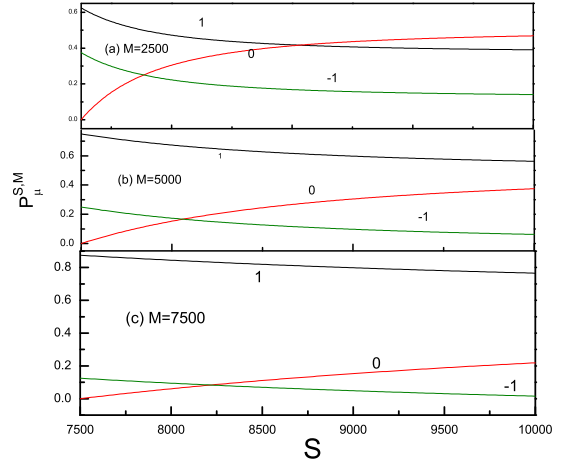}
\caption{\label{f1} \ $P_{\mu }^{S,M}$ as functions of $S$\ when
$N=10000$\ and $M$\ are given at 2500\ (a), 5000 (b), and 7500 (c).
$\mu $ is marked by the curve. }
\end{figure}

The $P_{\mu }^{S,M}$ with $N=10000,\ M=N/4,\ N/2,$ and $3N/4$\ \ are
plotted in Fig.\ref{f1}, the curves are not sensitive to $N,$ as it
appears in (\ref{e7a}-\ref{e7c}). For excited states, the total
spin-states are not necessary to be completely symmetric (e.g., for
the first excited band, both the spatial states and total
spin-states have the $\{N-1,1\}$ symmetry. \cite{bao05} However, if
the temperature is sufficiently low, only low-lying states are
concerned, where only a very small part of particles are excited. It
implies that most particles are condensed, and the spin-states of
these majority must be completely symmetric, while the effect of the
excited particles on the average spin-orientation is very small.
Thus the probability $P^{S,M}_{\mu}$ holds, in good approximation,
for all low-lying states.

When the final state is a thermodynamical distribution over the
eigen-states with the same $M$, the probability of a particle at
$\mu$ is
\begin{equation}
\mathbf{P}(M,\mu )=\Theta \sum_{i}e^{-E_{i}/T}P_{\mu }^{S_{i},\
M}\label{e13}
\end{equation}
where $i$ is the label of the levels, $E_{i}$ and $S_{i}$\ are the
corresponding energy and total spin, $\Theta $ is a constant just
for the normalization.

\begin{figure}
\includegraphics{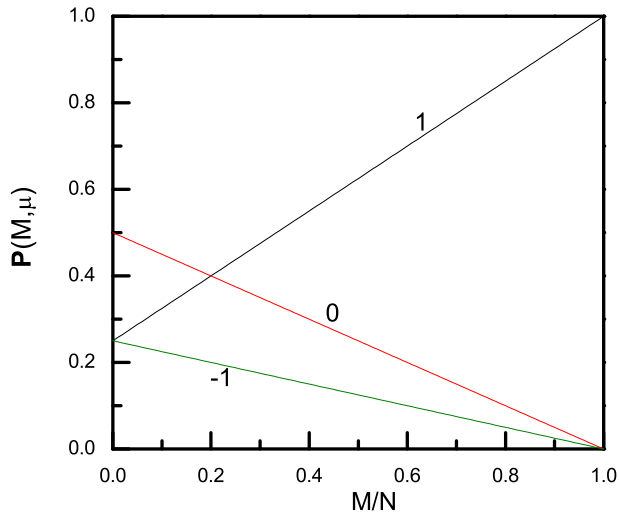}
\caption{\label{f2} $\mathbf{P}(M,\mu )$ as functions of $M/N$. $\mu
$ is marked by the curve.}
\end{figure}

From our previous study \cite{bao04,bao05,pang07} the low-lying
states of spinor condensates are divided into bands, the states in a
band have similar spatial wave functions but they are different in
$S$. The energy splitting inside the band is caused by the
spin-dependent atom-atom interaction. However, for realistic case,
the spin-dependence is weak. it was found from our calculation that
the splitting of energy levels in a band is very small. \cite{comm}
Hence, for the levels of a band the factor $e^{-E_{i}/kT}$ can be
roughly
considered as a constant, \cite{comm} thus the contribution of the $j$-th band to $%
\mathbf{P}(M,\mu )$ is just $\Theta _{j}\sum_{S}P_{\mu }^{S,\ M},$
where the summation of $S\ $is from $N$, $N-2,$to $M$ or $M+1$\
because $N-S$\ must be even, and $\Theta _{j}$ is a band-dependent
constant. Thus, when the contribution from all the bands are taken
into account, the normalized probability appears as
\begin{equation}
\mathbf{P}(M,\mu )=\frac{2}{N-M+2-\gamma} \sum_{S}P_{\mu }^{S,\
M}\label{e14}
\end{equation}
where $\gamma=0$ $(1)$ if $N-M$ is even (odd). $\mathbf{P}(M,\mu )$
is plotted in Fig.\ref{f2} , which is the probability of a boson at
$\mu $ if the condensate starts with $M$\ and finally arrives at
thermo-equilibrium. It is found that $\mathbf{P}(M,\mu )$ depends
only on $M/N$ as shown in Fig.\ref{f2}. Obviously,
due to (\ref{e7a},\ref{e7b},\ref{e7c}), $\mathbf{P}%
(-M,\mu )=\mathbf{P}(M,-\mu )$ . There are the following features.

(i) The Stern-Gerlach splitting of spinor condensates is described
by $\mathbf{P}(M,\mu )$, which is system-independent, i.e., it does
not depend on the species and the details of interactions, but is
essentially determined by symmetry.

(ii) Let $N_{\mu}$ be the number of particles at $\mu$ in the
initial state. $\mathbf{P}(M,\mu )$ also does not depend on the
details of $N_{mu}$ but only on $N_{1}-N_{-1}=M$\ . This coincides
with the experiment by Stenger, et al (Fig.2 of Ref.\cite{sten98}).

(iii) $\mathbf{P}(M,\mu )$ depends on $M/N$ nearly linearly.

(iv) When $M=0,\ \mathbf{P}(M,\mu )=1/4,\ 1/2,$ and 1/4 when $\mu
=1,0,$ and $-1$. \ This is also supported by the above experiment.

(v) When $M=N,\ \mathbf{P}(M,\mu )=1,0,$and $0$ if $\mu =1,0,$ and $%
-1 $ as expected.

(vi) When $M=N/2,\ \mathbf{P}(M,1)=5/8,\ \mathbf{P}(M,0)=2/8,$ and $%
\mathbf{P}(M,-1)=1/8$. \ This is supported by the experiment as
shown in Fig.5 of Ref.\cite{schm04}.

In fact, disregarding any set of initial $N_{1}$, $N_{0}$, and $%
N_{-1},$ the Stern-Gerlach splitting can be predicted based on
(\ref{e14}) or Fig.\ref{f2}.

In summary, it is recalled that, in explaining the Stern-Gerlach
splitting, the details of dynamics is not involved. Instead, the
assumption of arriving at thermo-equilibrium together with a strict
symmetry consideration play the role. Nonetheless, in the above
derivation, the total spin-state is assumed to be completely
symmetric. This is exactly true for the ground band and therefore
the above theory is rigorously valid at the low-temperature limit.
When the temperature is nonzero but is still low (say, $T\approx
100nk$) so that the number of excited particles is still small, the
above theory remains qualitatively valid.

Symmetry is well known to be crucial for various few-body systems.
This this paper gives an example that many-body systems are also
governed by symmetry.


\begin{thebibliography}{99}
\bibitem{comm} For an example, from \cite{bao05}, when $N=10000$,
the frequency of the harmonic trap $\omega =100/sec$, the band-width
$E_{bw}$ of the ground bands of $^{87}$Rb and $^{23}$Na condensates
are $3\times 10^{-9}$ and $1.2\times 10^{-8} meV$, respectively.
Accordingly, when $T=10^{-6}K$, the ratio of the probabilities of
staying in the highest and the lowest levels of the band
$e_{-E_{dw}}$ is 0.966 (Rb) or 0.871 (Na). When $T=10^{-3}K$, the
above two values become $0.99997$ (Rb) or $0.99986$ (Na).
\end{thebibliography}
\end{document}